\documentclass[%
reprint,
 amsmath,
 amssymb,
 aps,
 prl
]{revtex4-2}

\usepackage{graphicx}
\usepackage{subfig}
\usepackage{dcolumn}
\usepackage{bm}
\usepackage{caption}
\usepackage{balance}
\usepackage{bbm}
\usepackage{hyperref}
\hypersetup{
    colorlinks=true,
    linkcolor=red,
    filecolor=red,
    urlcolor=red,
    citecolor=blue, 
}

\usepackage[font=small,labelfont=bf]{caption}
\usepackage[symbol]{footmisc}
\usepackage{stackengine}

\newcommand\xrowht[2][0]{\addstackgap[.5\dimexpr#2\relax]{\vphantom{#1}}}

\newcommand{\bra}[1]{\left\langle #1 \right|}
\newcommand{\ket}[1]{\left|#1\right\rangle}
\newcommand{\braket}[2]{\left\langle#1 |  #2\right\rangle}

\newcommand{\braz}[1]{\langle #1 |}
\newcommand{\ketz}[1]{|#1\rangle}
\newcommand{\braketz}[2]{\langle#1 |  #2\rangle}

\let\f\frac
\let\p\partial

\begin{document}


\title{Quantum State Driving along Arbitrary Trajectories}

\author{\small Le Hu$^{1,2}$}
\email{lhu9@ur.rochester.edu}
\author{\small and Andrew N. Jordan$^{2,1}$}

\affiliation{$^1$Department of Physics and Astronomy, University of Rochester, Rochester, New York 14627, USA}
\affiliation{$^2$Institute for Quantum Studies, Chapman University, 1 University Drive, Orange, CA 92866, USA}

\date{\today}

\begin{abstract}
Starting with the quantum brachistochrone problem of the infinitesimal form, we solve the minimal time and corresponding time-dependent Hamiltonian to drive a pure quantum state with limited resources along arbitrary pre-assigned trajectories. It is also shown that out of all possible trajectories, with limited resources, which are physically accessible and which are not. The solution is then generalized to the mixed quantum state cases, and applied to trajectories parameterized by single or multiple parameters with discrete or continuous spectrum. We then compare the solution to that of the counterdiabatic driving, and show how the Berry phase is directly involved in both driving processes.
\end{abstract}

\maketitle


\section{I. Introduction}
An important problem in quantum control theory is how to act on a quantum system so as to drive the state to a desired goal. The idea of quantum speed limit, first rigorously derived in \cite{mandelstam1945energy}, and latter extensively developed by \cite{fleming1973unitarity,bhattacharyya1983quantum,margolus1998maximum, giovannetti2003quantum,deffner2013energy,giovannetti2004speed,bures1969extension,taddei2013quantum,caneva2009optimal,jones2010geometric,deffner2013energy,zhang2014quantum,hegerfeldt2013driving,funo2019speed,liu2015quantum,il2021quantum,wu2018quantum}, denotes the minimal time $\tau_\text{QSL}$ it takes for an undriven pure quantum state $\ket{\psi_i}$ to evolve to $\ket{\psi_f}$. It turns out that for the discrete spectrum case in a closed quantum system, $\tau_\text{QSL}$ is given by the larger value \cite{deffner2013energy} of the Mandelstam-Tamm (MT) bound \cite{mandelstam1945energy} and the Margolus-Levitin (ML) bound \cite{margolus1998maximum},
\begin{equation} \label{eq1}
	\tau_\text{QSL} = \max \{  \underbrace{\frac{\hbar \arccos{|\braket{\psi_i}{\psi_f}|}}{\Delta E}}_\text{MT bound}, \underbrace{\frac{\hbar \arccos{|\braket{\psi_i}{\psi_f}|}}{E}}_\text{ML bound}\},
\end{equation}
where $\Delta E=(\langle H\rangle^{2}-\left\langle H^{2}\right\rangle)^{1 / 2}$ and $E=\langle H\rangle-E_\text{ground}$. Generalizations to the mixed quantum state \cite{deffner2013energy}, thermal state \cite{il2021quantum}, and open quantum systems \cite{deffner2013quantum, taddei2013quantum, zhang2014quantum, funo2019speed} have also been performed.

However, if one is allowed to manipulate the Hamiltonian $H(t)$ during the evolution $\ket{\psi_i} \to \ket{\psi_f}$ to minimize the evolution time, then it becomes a time-optimization problem known as quantum brachistochrone, in analogy to the famous brachistochrone problem posed by Bernoulli in 1696. The quantum brachistochrone problem was first examined by Carlini \textit{it al.}\cite{carlini2006time} In their paper, they solved for the optimal Hamiltonian to drive a given initial state $\ket{\psi_i}$ to a given final state $\ket{\psi_f}$ in a time-optimal way, through the method of the Lagrangian multiplier and variational principle. It is found that with limited resources, meaning specifically that the the variance of the Hamiltonian $\Delta H^2$ is bounded by some value $\omega^2$ and no other constraints, the optimal Hamiltonian, which is time-independent, is given by (we set $\hbar =1$)

 \begin{figure}[!htp]
\subfloat{\includegraphics[width=.4\textwidth,scale=1]{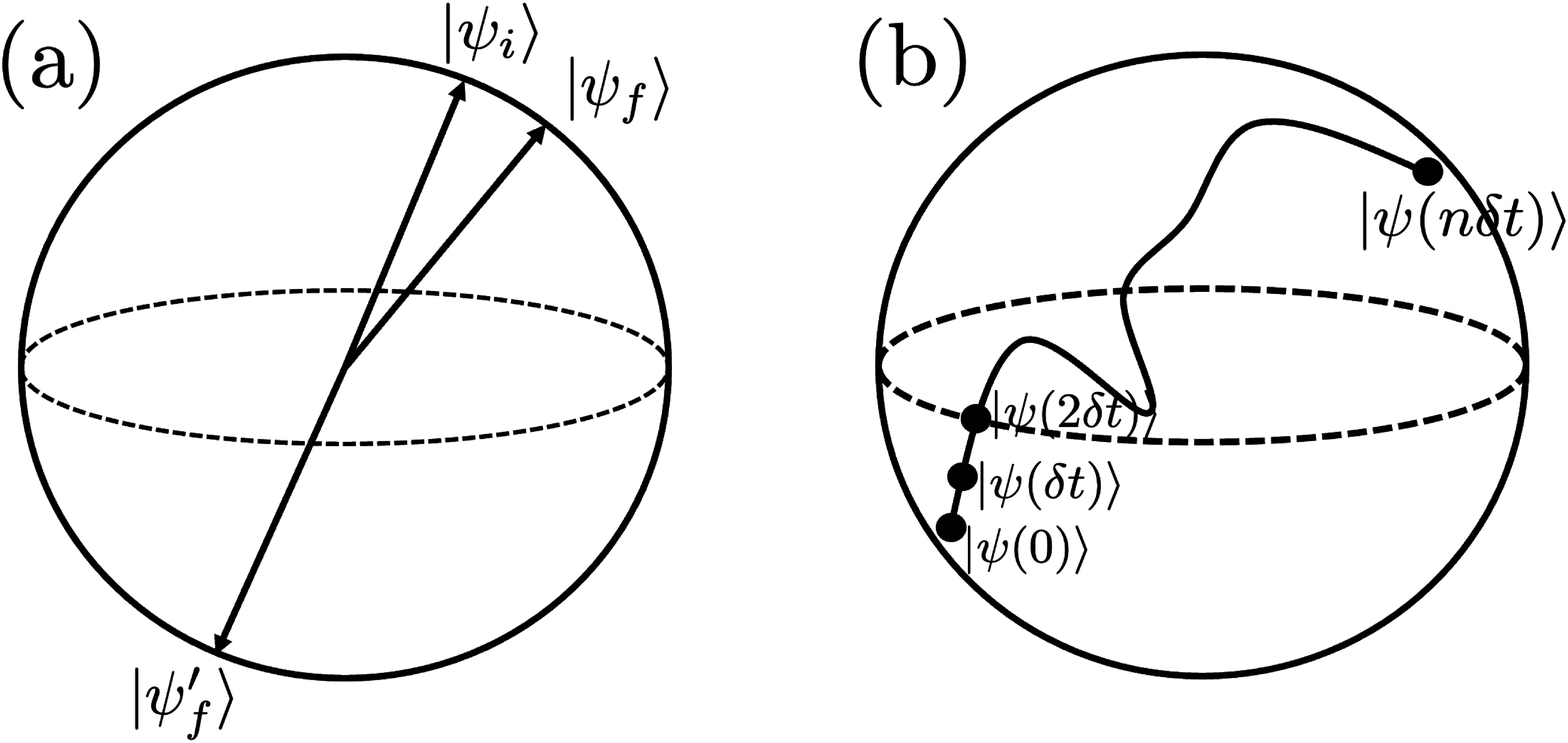}\hfill}
\caption{\label{fig:epsart1} (a) State $\ketz{\psi^\prime_f}$ is constructed by the Gram-Schmidt process from $\ket{\psi_i}$ and $\ket{\psi_f}$ such that $\braketz{\psi_i}{\psi^\prime_f}=0.$ As visualized by the Bloch sphere for two-level system, this means for $\ket{\psi_i}$ to time-optimally evolve to $\ket{\psi_f}$, it is sufficient to let it evolve along the great circle formed by $\ket{\psi_i} \to \ket{\psi_f} \to \ketz{\psi^\prime_f}$. The idea also applies to the $n$-level quantum system. (b) For any continuous quantum trajectories $\ket{\psi(t)}$, we can always cut it to an infinite number of infinitesimal pieces, then apply the time-optimal evolution strategy shown in panel (a) for each pieces. 
} 
\end{figure}

 \begin{equation}\label{eq2}
 	H=i \omega (\ketz{\psi^\prime_f}\bra{\psi_i}-\ketz{\psi_i}\braz{\psi^\prime_f}),
 \end{equation}
 where $\ketz{\psi^\prime_f}$ is the Gram-Schmidt orthonormalized state with respect to $\ket{\psi_i}$ such that $\braketz{\psi^\prime_f}{\psi_i}=0$; see Fig.\,\ref{fig:epsart1}(a) for a more intuitive understanding. The optimal time then is given by
 \begin{equation}\label{eq3}
 	T=\frac{1}{\omega}\arccos |\braket{\psi_i}{\psi_f}|.
 \end{equation}

\noindent These concepts are limited to optimizing the time taken between fixed initial and final conditions.
 
\section{II. Arbitrary Quantum Trajectories}
Here, we give the problem to be solved in this paper. Suppose that instead of studying evolutions between two fixed endpoints, one with limited resources is interested in driving time-optimally a quantum state  along pre-assigned continuous trajectories $\ket{\psi(t)}$, then what Hamiltonian $H(t)$ shall one need? Since we already know the optimal Hamiltonian that drives $\ket{\psi_i}$ to $\ket{\psi_f}$, we can cut the trajectories into infinite number of infinitesimal pieces, with $\ket{\psi(t)}$ being the initial state and $\ket{\psi(t+\delta t)}$ being the final state, so as to apply the piecewise constant optimal Hamiltonian given by Eq.\,(\ref{eq2}) (Fig.\,\ref{fig:epsart1}(b))
 \begin{equation}
 H(t)=i\omega(t)(\ket{\psi^\prime (t+\delta t)}\bra{\psi(t)}-\ket{\psi(t)}\bra{\psi^\prime (t+\delta t)}).
 \end{equation}
By using Taylor series,
 \begin{equation}
 	\begin{aligned}
 		|\psi(t+\delta t)\rangle&=|\psi(t)\rangle+\delta t\left|\partial_{t} \psi(t)\right\rangle+\mathcal{O}(\delta t^2)\\
 		\partial_{t}|\psi(t+\delta t)\rangle&=\left|\partial_{t} \psi(t)\right\rangle+\delta t\left|\partial_{t t} \psi(t)\right\rangle+\mathcal{O}(\delta t^2),\\
 	\end{aligned}
 \end{equation}
 and constructing $\ket{\psi^\prime(t)}$ by the Gram-Schmidt process from $\ket{\psi(t-\delta t)}$ and $\ket{\psi(t)}$,
 \begin{equation}
 	\begin{aligned}
 		\left|\psi^{\prime}(t)\right\rangle&=\frac{|\psi(t)\rangle-\langle\psi(t-\delta t)|\psi(t)\rangle|\psi(t-\delta t)\rangle}{\sin \Omega(t-\delta t)}\\
 		\sin \Omega(t)&=\sqrt{1-|\langle\psi(t)|\psi(t+\delta t)\rangle|^{2}},
 	\end{aligned}
 \end{equation}
 where $\sin \Omega(t)$ serves as a normalization factor, one may rewrite $\ket{\psi^\prime{(t+\delta t)}}$ in terms of $\ket{\psi(t)}$, $\ket{\partial_t\psi(t)}$ and $\ket{\partial_{tt}\psi(t)}$. Dropping the higher-order terms $\delta t^2$, and rewriting outstanding $\delta t$ according to Eq.\,(\ref{eq3}),
 one obtains
 \begin{equation}
 	H(t) = i\left(\left|\partial_{t} \psi(t)\right\rangle\langle\psi(t)|-| \psi(t)\rangle\left\langle\partial_{t} \psi(t)\right|\right).
 \end{equation}
By requiring $H(t)$ to satisfy the Schr\"{o}dinger equation $i\ket{\p_t \psi(t)}=H(t)\ket{\psi(t)}$, we need to have $\braket{\p_t \psi(t)}{\psi(t)}=0$. This can be achieved via fixing the $U(1)$ gauge by defining 
 \begin{equation} \label{eq5}
 \ketz{\tilde\psi(t)}=e^{i\phi(t)} \ket{\psi(t)},
  \end{equation}
   such that $\braketz{\p_t \tilde{\psi}(t)}{\tilde{\psi}(t)}=0,$ which gives
 \begin{equation} \label{eq6}
 	\phi(t)=\int -i \braket{\p_t \psi(t)}{\psi(t)}dt \quad \in \mathbb{R}.
 \end{equation}
 This is essentially canceling the open-path Berry phase, as $t$ can also be regarded as a parameter for the purpose of calculating the Berry phase. One then obtains the optimal Hamiltonian
  \begin{equation} \label{eq7}
  \begin{aligned}
 	H(t) &= i(|\partial_{t} \tilde{\psi}(t)\rangle\langle\tilde{\psi}(t)|-| \tilde{\psi}(t)\rangle\langle\partial_{t} \tilde{\psi}(t)|)+\dot{\phi}(t) \mathbbm{1},
 	  \end{aligned}
 \end{equation} 
 which is one of our main results. Note that this is equivalent to solving the Schr\"{o}dinger equation in a reverse way.
 The $\dot{\phi}(t)$ term is used to cancel the $e^{i\phi(t)}$ phase term such that $H(t)$ will actually drive $\ket{\psi(t)}$, not $\ketz{\tilde{\psi}(t)}$. Other selections of real functions can also be added to $H(t)$ as long as the $e^{i\phi(t)}$ phase term is canceled.
One can thereby drive any continuously changing pure quantum state $\ket{\psi(t)}$, at least in principle, by applying a control Hamiltonian $H_\text{c}(t)$ given by $H_\text{c}(t)=H(t)-H_\text{s}(t)$, where $H_\text{s}(t)$ is the Hamiltonian of the original system. Such a control Hamiltonian can be implemented via various techniques, e.g. linear combinations of unitary operators (LCU)\cite{childs2012hamiltonian}, truncated Taylor series (BCCKS)\cite{PhysRevLett.114.090502}, qubitization\cite{low2019hamiltonian}, unitary decomposition of operators \cite{PhysRevLett.127.270503}, etc.
 
 It is easy to check that
 \begin{equation} \label{eq8}
 \begin{aligned}
 	\langle H(t) \rangle_t&=\bra{\psi(t)}H(t)\ket{\psi(t)}=\dot{\phi}(t)\\
 	(\Delta H(t))^2 &=\braket{\partial_t \psi(t)}{\partial_t \psi(t)}-\dot{\phi}^2(t)\equiv \omega^2(t),
  \end{aligned}
 \end{equation}
 which means if our resources are limited in a way that
 \begin{equation}
 	\sup_t {(\Delta H(t))^2} = \omega_\text{max}^2,
 \end{equation}
 with no other constraints, then only the paths satisfying
 \begin{equation}
 	\lVert\ket{\partial_t \psi(t)}\rVert \leq \sqrt{\omega^2_\text{max}+\dot{\phi}^2(t)}
 \end{equation}
 are physically accessible. It is then obvious to see that the minimal time is achieved when the equality holds for all time. Since $\ket{\psi(t)}$ and $\ketz{\tilde{\psi}(t)}$ are essentially the same quantum state up to a global $U(1)$ phase, we will consider $\ketz{\tilde{\psi}(t)}$ only in the following, indicating $\dot{\phi}(t)=0$. Readers can always transform back and forth between $\ket{\psi(t)}$ and  $\ketz{\tilde{\psi}(t)}$ according to $\ketz{\tilde\psi(t)}=e^{i\phi(t)} \ket{\psi(t)}$ and Eq.\,(\ref{eq6}).

 For a general $n$-level system expanded in its eigenbasis $\ketz{\tilde{\psi}(t)}=\sum_i^n a_i(t)\ket{\psi_n}$, where $a_i(t)$ is some function such that $\sum_i|a_i(t)|^2=1$, the requirement $\lVert\ketz{\partial_t \tilde{\psi}(t)}\rVert = \omega_\text{max}$ is equivalent to $\sqrt{\sum_{i}^{n}|\dot{a}_{i}(t)|^2}=\omega_\text{max}$
where the dot over $\dot{a}$ denotes the time derivative. It is very interesting to see how it enjoys the similar form as that of a classical free particle, parameterized by $\{x_i(t)\}$ in the position space and with speed $v_\text{max}$, traveling in $n$-dimensional space
$\sqrt{\sum_{i}^{n} \dot{x}^{2}_{i}(t)}=v_{\max },$ except that in general $a_i(t) \in \mathbb{C}$ whereas $x_i(t) \in \mathbb{R}$.

Now to make the problem a little bit more realistic, let us suppose we want to drive the quantum state along a preassigned $s$-parametrized trajectory $\ketz{\psi(x(s))}$, which can be transformed to its $U(1)$ gauge equivalent state $\ketz{\tilde{\psi}(x(s))}=e^{i \phi(x(s))}\ketz{\psi(x(s))}$ such that $d\phi/dx=0$, where $\phi(x(s))=\int -i \braket{\p_x \psi(x(s))}{\psi(x(s))}(dx/ds)~ds$. For now, we do not know the explicit time-dependency of the general parameter $x$. Here, the general parameter $x$ can denote position, momentum, the angular frequency of an oscillator, or any other parameters we want to manipulate. We would like to solve for time $t$ in terms of a function of $x(s)$ such that $t=f(x(s))$ to obtain $\ketz{\tilde{\psi} (f^{-1}(t))}$, so that we can apply all the previous results we just derived. This can be performed by
\begin{equation} \label{eq15}
\begin{aligned}
	t(x_{s_0 \to s_1})&
	 =\int_{t_0}^{t_1} \frac{\sqrt{\omega^2(t)}}{\sqrt{\omega^2(t)}}~dt=\int_{s_0}^{s_1} \frac{\lVert \ketz{\partial_x \tilde{\psi}(x(s^\prime))\f{dx}{ds^\prime}}\rVert}{\omega(x(s^\prime))}~ds^\prime,
\end{aligned}
\end{equation}
where we have used Eq.\,(\ref{eq8}) and the change in variables. The above equation of $t(x_{s_0 \to s_1})$ denotes the time needed to time-optimally drive the quantum state along the path $x(s_0) \to x(s_1)$ given bounded energy variance $\omega^2(t)$.

\begin{figure}[!tp]
\subfloat{\includegraphics[width=0.35\textwidth,scale=1]{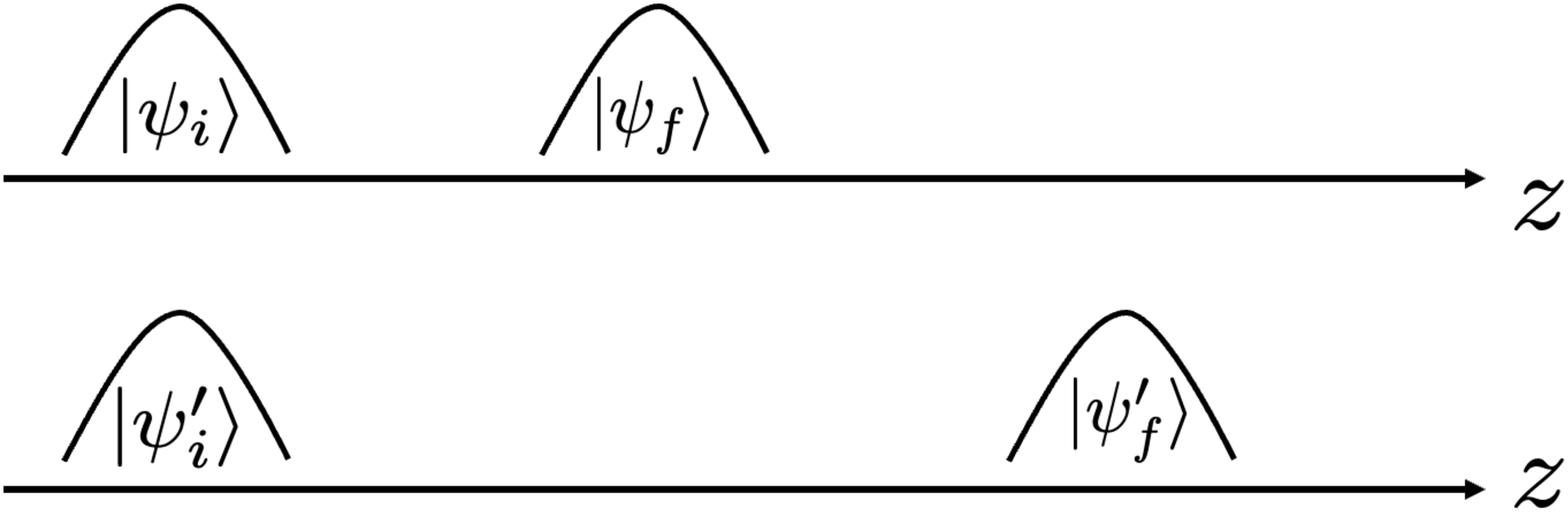}\hfill}
\caption{\label{fig:epsart2} Schematic of initial and final wave packets under different spatial separations in the position space. The formula $\arccos{|\braket{\psi_i}{\psi_f}|}$ to characterize the distance between two quantum states clearly is not suitable to the case of continuous spectrum. One obtains $\arccos|\int\psi_f^*(z)\psi_i(z) dz| \approx \arccos|\int{\psi_f^\prime}^*(z)\psi^\prime_i(z) dz| \approx\pi/2$, but obviously the two wave packets in the lower figure are much more separated, indicating the breakdown of MT/ML formulas.}
\end{figure}

The above idea, applicable to the discrete-spectrum and single-parameter case, can also be easily generalized to the continuous-spectrum (e.g. position or momentum) and multi-parameter case. One then may utilize the generalized formula to the continuous-spectrum case to calculate corresponding quantum speed limit, which cannot be performed by the original MT or ML formula (Eq.\,(\ref{eq1})). See Fig. \ref{fig:epsart2} and Table \ref{tab:1} for details.

\begin{table}[h] 

  \centering
  \begin{tabular}{lcr}
\textbf{\quad \quad \quad Discrete} & \textbf{Continuous ($n$-dimensional)} \\  \hline \xrowht{40pt}
$\displaystyle t=\int \frac{\|\ketz{\partial_{x} \tilde{\psi}(x(s))}\f{dx}{ds}\|}{\omega{(x(s)) }} ds,$      &    $\displaystyle t=\int \frac{\sqrt{\int| \f{\p\tilde{\Psi}(x, \vec{z})}{\p x}\f{d x}{d s}|^{2} d^n z}}{\omega{(x(s)) }}ds$     \\  \hline \xrowht{40pt}
$\displaystyle t=\int \frac{\||\partial_{\vec{x}} \tilde{\psi}(\vec{x}(s)) \frac{d \vec{x}}{ds}\rangle \|}{\omega(\vec{x}(s))}ds,$         & $\displaystyle t=\int \frac{\sqrt{\int|\frac{\partial \tilde{\Psi}(\vec{x}(s), \vec{z})}{\partial \vec{x}} \frac{d \vec{x}}{ds}|^{2} d^n z}}{\omega{(\vec{x}(s)) }}ds$  \\     
  \end{tabular}
  \caption{The time $t$ needed to drive the quantum state along the trajectory parametrized by the generalized parameter $\vec{x}(s)=(x_1(s), x_2(s), \cdots, x_i(s))^T$ for the discrete/continuous spectrum and single/multiple-parameter cases. Note that $\vec{x}$ can always be parametrized by another parameter $s$ so the multiple-parameter case can be reduced to the single one. The $\vec{z}$ in the equations on the right denotes the variable one needs to integrate over the infinite-dimensional Hilbert space, e.g., the position or momentum. The symbol $\tilde{\Psi}(\vec{x},\vec{z})$ denotes the wave function in $L^2(\mathbb{R}^n)$ such that $\ketz{\tilde{\psi}(\vec{x}, \vec{z})}=\int \tilde{\Psi}(\vec{x},\vec{z}) \ket{\vec{z}}d^nz$.
}   \label{tab:1}
\end{table}

\section{III. The connection to counterdiabatic driving}
Readers familiar with counterdiabatic driving, which has many applications in quantum metrology \cite{yang2017quantum, naghiloo2017achieving, jordan2018quantum,cabedo2020shortcut,yang2022variational}, may have noted some interesting similarities between the optimal Hamiltonian we just derived and the counterdiabatic Hamiltonian appears in the literature \cite{berry2009transitionless, del2013shortcuts,torrontegui2013shortcuts,guery2019shortcuts},
 \begin{equation}
 	H_\text{CD}(t)=-H_s(t)+i\sum_n \ket{\partial_t \psi_n(t)}\bra{\psi_n(t)},
 \end{equation}
   where $H_s(t)$ is the original Hamiltonian of the system and $\ket{\psi_n(t)}$ is the instantaneous eigenstate of $H_s(t)$. To see the connection more clearly and help readers less familiar with counterdiabatic driving to understand, let us go back the the adiabatic theorem. The adiabatic theorem states that if a quantum state $\ket{\psi_n(0)}$ is at the instantaneous eigenstate of a slowly changing Hamiltonian, then it will continue to be the same instantaneous eigenstate at a later time $t$ up to a phase $e^{i\theta_n(t)}e^{i\gamma_n(t)}$, where $\theta_n(t)=-\int_0^t E_n(t^\prime) dt^\prime$ is the dynamic phase and $\gamma_n(t)= \int_0^t i\braketz{\psi_n(t^\prime)}{\dot{\psi}_n(t^\prime)}dt^\prime$ is the Berry phase. 
 
Although the physical meaning of the dynamic phase is very clear, as $E_n(t)$ is nothing but the instantaneous eigenenergy, the physical meaning of the Berry phase, especially the $i\braketz{{\psi}_n(t)}{\dot{\psi}_n(t)}$ term, is not so clear. To give further insight, we note that if $\ket{\psi(t)}$ satisfies the Schr\"{o}dinger equation, the form of $i\braketz{{\psi}(t)}{\dot{\psi}(t)}$ is equivalent to $\langle H(t)\rangle_t$ if one restores $i\partial_t$ to $H(t)$,
 \begin{equation}
 	i\langle\psi(t)|\dot{\psi}(t)\rangle=\langle\psi(t)|i \dot{\psi}(t)\rangle=\langle\psi(t)|H(t)| \psi(t)\rangle=\langle H(t)\rangle_t,
 \end{equation}
indicating $i\braketz{{\psi(t)}}{\dot{\psi}(t)}$ has a clear physical meaning. We would like to apply the same trick to $i\braketz{\psi_n(t)}{\dot{\psi}_n(t)}$ but unfortunately, the instantaneous eigenstate does generally not satisfy the Schr\"{o}dinger equation: $i \partial_t \ket{\psi_n(t)} \neq H(t) \ket{\psi_n(t)}$. To get around this, let us suppose for the moment there exists such Hamiltonian, denoted as $H^\prime(t)$, which happens to satisfy the Schr\"{o}dinger equation for each of its instantaneous eigenstates
\begin{equation}
	i \partial_t \ket{\psi_n(t)} = H^\prime(t) \ket{\psi_n(t)} \quad n=0,1,2,\dots,
\end{equation}
then according to the same argument, $i\braketz{{\psi}_n(t)}{\dot{\psi}_n(t)}$ is equivalent to the expectation value of $H^\prime(t)$.
 \begin{equation}
 	i\langle\psi_n(t)|\dot{\psi}_n(t)\rangle=\langle\psi_n(t)|H^\prime(t)| \psi_n(t)\rangle=\langle H^\prime(t)\rangle_t.
 \end{equation}
 The Berry phase can thereby be interpreted as the time integral of the expectation value of $H^\prime(t)$, i.e., $ 	\gamma_n(t)=\int_0^t \langle H^\prime(t^\prime)\rangle dt^\prime.$
 Hence the total phase can be rewritten as
 \begin{equation}
 	e^{i \theta_n(t)}e^{i \gamma_n(t)} = e^{\int_0^t-i \langle H_s(t^\prime)-H^\prime(t^\prime)\rangle dt^\prime}
 \end{equation}
 Fortunately, the explicit expression of $H^\prime(t)$ is not difficult to find
 \begin{equation} \label{eq20}
 	H^\prime(t)=i\sum_n \ket{\partial_t\psi_n(t)}\bra{\psi_n(t)}.
 \end{equation}
 Note that, even though only the term $i\ket{\p_t \psi_m(t)}\bra{\psi_m(t)}$ is really driving the quantum evolution of $\ket{\psi_m(t)}$, all other $\dim H^\prime-1$ terms should still be included in order to keep the general Hermiticity of $H^\prime(t)$.
 
 The benefit of the above interpretation is that if one wants to get rid of the phase $e^{i \theta_n(t)}e^{i \gamma_n(t)}$, one only needs to apply a control Hamiltonian, also called the counterdiabatic Hamiltonian $H_\text{CD}(t)$ in this case, by letting $H_\text{CD}(t)$ just be the negative of $H_s(t)-H^\prime(t)$,
 \begin{equation}
 	H_\text{CD}(t)=-(H_s(t)-H^\prime(t)).
 \end{equation}
 The total Hamiltonian that actually guides the evolution of the system, with $H_\text{CD}(t)$ applied, is then
 \begin{equation}
 H_\text{total}(t) = H_s(t) +H_\text{CD}(t) = H^\prime(t).
  \end{equation}
 That is to say, the ``nice'' Hamiltonian $H^\prime(t)$ we just proposed is not just hypothetical, but a real one, which actually governs the system, if $H_\text{CD}(t)$ is applied. 
 
Compare the above $H^\prime(t)$ with the arbitrary-trajectory-driving Hamiltonian we derived earlier
  $$H^{\prime\prime}(t)=i(|\partial_{t} \tilde{\psi}(t)\rangle\langle\tilde{\psi}(t)|-| \tilde{\psi}(t)\rangle\langle\partial_{t} \tilde{\psi}(t)|)+\dot{\phi}(t) \mathbbm{1}.$$
  where we have used double primes to clearly distinguish it from $H^\prime(t)$. What are their connections? The idea is that, if $H^{\prime\prime}(t)$ can guide any trajectories, it must also be able to guide counterdiabatic driving evolution, i.e., $H^{\prime\prime}(t)$ should be able to play the role of $H^\prime(t)$. On the contrary, any trajectories $\ket{\psi(t)}$ a quantum state undergoes can be viewed as being at the instantaneous eigenstate $\ket{\Psi_m(t)}$ of some unknown Hamiltonian $\mathbbm{H}(t)$. Hence one will always be able to apply the counterdiabatic driving $\mathbbm{H}_\text{CD}(t)$ such that $H^\prime(t)=\mathbbm{H}(t)+\mathbbm{H}_\text{CD}(t)=i(\sum_n\ket{\partial_t \Psi_n(t)}\bra{\Psi_n(t)})$, which does not need the $\dot{\phi}(t)$ term because it is already implicitly included. In this case, one needs to ``make up'' the other $\dim \mathbbm{H}-1$ terms orthonormal to $\ket{\Psi_m(t)}$. Those terms do not really help the driving of the desired state $\ket{\Psi_m(t)}$, but are merely used as a means to keep the general Hermiticity of $H^\prime(t)$, therefore this approach is less convenient when $\dim \mathbbm{H}$ becomes large (and impossible if $\dim \mathbbm{H} \to \infty$). To summarize, both $H^{\prime}(t)$ and $H^{\prime\prime}(t)$ contain the key driving term $i\ket{\p_t \psi(t)}\bra{\psi(t)}$, and some other ``useless'' terms merely to keep the Hermiticity.
  
  \textit{The generalization to mixed quantum state.}---Below, we briefly describe how one can generalize the above results to the mixed quantum state cases. Let the pre-assigned density matrix trajectories $\rho(t)= \sum_n p_n \ket{n(t)}\bra{n(t)}$ at time $t$ be diagonalized in its instantaneous eigenvector $\ket{n(t)}$ basis with $\sum_n p_n =1$.
We can obtain $\ket{\tilde{n}(t)}=e^{i\phi_n(t)}\ket{n(t)}$ by Eq.\,(\ref{eq6}) for each of the eigenvectors in $\{\ket{n(t)}\}$, from which we can define $\tilde{\rho}(t)=\sum_n p_n \ketz{\tilde{n}(t)}\bra{\tilde{n}(t)}$. The Hamiltonian $\tilde{H}(t)$ driving $\tilde{\rho}(t)$ is then given by
  \begin{equation}\label{eq23}
  \tilde{H}(t)=\f{i}{2}\sum_n (\ketz{\dot{\tilde{n}}(t)}\braz{\tilde{n}(t)}-\ketz{\tilde{n}(t)}\braz{\dot{\tilde{n}}(t)}).
    \end{equation}
  This can be verified by taking the time derivative of $\mathbbm{1}=\sum_n \ket{\tilde{n}(t)}\bra{\tilde{n}(t)}$ to obtain $\sum_n \ket{\dot{\tilde{n}}(t)}\bra{\tilde{n}(t)}=-\sum_n \ket{\tilde{n}(t)}\bra{\dot{\tilde{n}}(t)}$, transforming the Hamiltonian into $H(t)=i\sum_n \ket{\dot{\tilde{n}}(t)}\bra{\tilde{n}(t)}$, which interestingly is exactly of the form Eq.\,(\ref{eq20}). This suggests a canonical way of applying Eq.\,(\ref{eq20}) to drive pure quantum states along designed trajectories, since every pure quantum state can be written in the density matrix form. The energy variance is given by
  \begin{equation}
  	(\Delta H(t))^2=\sum_{n, m} p_m |\langle \tilde{m}(t)| \dot{\tilde{n}}(t)\rangle|^2.
  \end{equation}
   By a similar argument, if the density state $\tilde{\rho}(x(s))$ is parametrized by a general single parameter $x$ for which we need to solve explicitly the dependency on $t$, then for the discrete spectrum case, it is given by
  \begin{equation}
  \begin{aligned}
  	t(x_{s_0 \to s_1})&= \int_{s_0}^{s_1} \frac{ \sqrt{\sum_{n,m} p_m |\braketz{\tilde{m}(x(s))}{\partial_x \tilde{n}(x)\f{\p x}{\p s}}|^2}}{\omega(x(s))} ds.\\
  	  \end{aligned}
  \end{equation}
  The expression for multi-parameter and continuous-spectrum case can be similarly derived.

\section{IV. Example}
\subsection{A. Landau-Zener Model}
To show explicitly how the results we obtained can be made use of, consider the Landau-Zener model with Hamiltonian $H_\text{LZ}(\Gamma)$,
  \begin{equation}
  	H_\text{LZ}(\Gamma(t))=\begin{pmatrix}
  		\Gamma(t)&\epsilon\\
  		\epsilon&-\Gamma(t)
  	\end{pmatrix}.
  \end{equation}
  At the current stage, the only thing we know is that $\Gamma$ will begin changing monotonically at time $t=-T$ and stop changing at some final time $t=T$ where $T$ is not yet determined, but under the constraint $\Gamma(-T)=-\Gamma(T)=-\Gamma_0$ such that the Hamiltonian has symmetric endpoints. We want our quantum state, starting at the instantaneous ground state $\ket{\psi_-(\Gamma(-T))}$ of $H_\text{LZ}(\Gamma(-T))$, to evolve along the instantaneous ground state of $H_\text{LZ}(\Gamma)$ as it changes. We may apply \textit{any} control Hamiltonian to accelerate the process, but the total Hamiltonian is subject to the constraint $\Delta H_\text{}(t)^2 \leq \omega^2_\text{max}$ due to limited resources.
  
  For simplicity, let us consider $t \geq 0$ only as $t \leq 0$ can be similarly obtained. 
  To calculate the minimal time it takes for $\ket{\psi_-(0)} \to$  $\ket{\psi_-(\Gamma_0)}$, we only need to apply Eq.\,(\ref{eq15}) and let $\omega(\Gamma)=\omega_\text{max}$ for all $\Gamma$,
  \begin{equation}
\begin{aligned}
  	t(\Gamma)&=\int_{0}^{\Gamma}\frac{\|\ket{\partial_{\Gamma^\prime} \psi_-(\Gamma^\prime)} \|}{{\omega_\text{max}} } |d\Gamma^\prime|=\f{\arctan(\Gamma/\epsilon)}{2\omega_\text{max}}
  \end{aligned}
  \end{equation}
  where we have used the fact that $\ket{\psi(\Gamma)}=\ketz{\tilde{\psi}(\Gamma)}$. Under the limit $\Gamma_0 \to \infty$, we have $\lim_{\Gamma_0 \to \infty}2t(\Gamma_0)=\frac{\pi/2}{\omega_\text{max}},$
  which is exactly the quantum speed limit obtained by Eq.\,(\ref{eq1}); see also Refs.\,\cite{bason2012high,hegerfeldt2013driving,poggi2013quantum}. The expression of $t(\Gamma)$ gives $\Gamma(t)=\epsilon\tan{(2t\omega_\text{max})}$, from which we can obtain the expression of $\ket{\psi_-(\Gamma(t))}$ immediately. Apply Eq.\,(\ref{eq7}) to obtain the optimal total Hamiltonian $H$ for $t\geq0$ (and via the similar way for $t\leq0$),
  \begin{equation}
  	H(t) =\begin{pmatrix}
  		0 & i\omega_\text{max}\\
  		-i \omega_\text{max} &0
  	\end{pmatrix},\quad t \in [-t(\Gamma_0),t(\Gamma_0)]
  \end{equation}
  Note that the Hamiltonian is different from the results in literature \cite{bason2012high,hegerfeldt2013driving,poggi2013quantum} where only up to two operators, $\sigma_x$ and $\sigma_z$, are allowed for manipulation.
  
\subsection{B. Moving squeezing Gaussian wave packet}

The ground-state wave function centered at $z=-x$ of the 1-$d$ quantum simple harmonic oscillator is a Gaussian function
  \begin{equation}
  \psi_0((x, \omega);z)=\left(\frac{m\omega}{\pi \hbar}\right)^{1/4}\exp{\left(-\frac{m \omega}{2 \hbar}(z+x)^2\right)},
    \end{equation}
   where $x$ and $\omega$ are the tunable parameters and $z$ denotes the position.
  Suppose we want the Gaussian wave packet to travel along the $+z$-direction, whereas, increasing the angular frequency $\omega$ of the oscillator, such that the parameters $x$ and $\omega$ are satisfied by the following parametric equation
  \begin{equation}
  	\left\{\begin{array}{l}x(s)=\mu s \\ \omega(s)=\omega_0/ s^2,\end{array}\right.
  \end{equation}
  with dimensionless parameter $s$ decreasing from 1 to $s_f$ where $s_f \in (0,1)$, and $\mu$ is a factor with unit $\text{m}$ such that $\mu s$ also has unit $\text{m}$. By applying a control Hamiltonian and under limited resources, i.e. $\int |\dot{\psi}_0((x(t),\omega(t));z)|^2 dz= \epsilon^2\hbar^2=const.,$
 \begin{figure}[!tbp]
\subfloat{\includegraphics[width=.4\textwidth,scale=1]{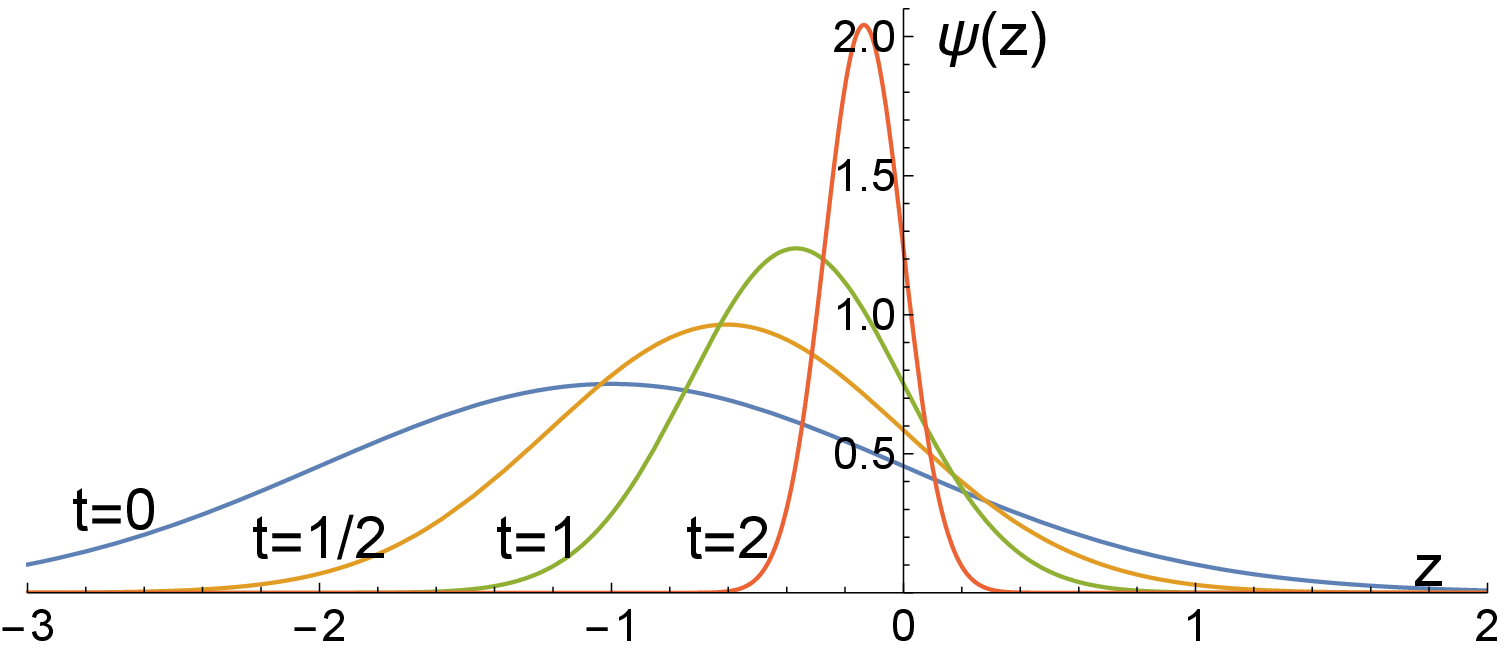}\hfill}
\caption{\label{fig:epsart3} The wave function $\psi_0(z(t))$ at different time $t$. We set $m=\omega_0=\mu=\hbar=1.$
} 
\end{figure}
  we want the $s$-parameterized wave packet $\psi_0((x(1),\omega(1));z)$ evolves to its final state $\psi_0((x(s_f),\omega(s_f));z)$ time-optimally. To solve the problem, we apply Eq.\,(\ref{eq15}) (remember to put $\hbar$ back), and use the fact that $\psi_0(s)=\tilde{\psi}_0(s)$,
  \begin{equation}
  \begin{aligned}
  	&t(s_f)=\int_{1}^{s_f} \frac{\sqrt{\int_{-\infty}^{\infty} |\partial_s \psi_0((x(s),\omega(s));z)|^2\hbar^2dz}}{\epsilon\hbar}|ds|\\
  	&\Rightarrow s_f(t)=e^{-\eta \epsilon t} \quad \text{,} \quad \text{$t \in [0, t(s_f)]$}
  \end{aligned}
  \end{equation}
where $\eta=\sqrt{2\hbar/(\mu^2 m\omega+\hbar)}$. Insert $s_f(t)$ back to $\psi_0((x(s),\omega(s));z)$ to obtain the wave function (Fig. \ref{fig:epsart3})
\begin{equation} \label{eq32}
	\psi_0(z,t)=\left( \frac{m \omega_0 }{\pi \hbar e^{-2\eta \epsilon t}} \right)^{1/4}\exp\left(-\frac{m\omega_0}{2\hbar e^{-2 \eta\epsilon t}} (z+\mu e^{-\eta \epsilon t})^2\right)
\end{equation}

One can thereby apply Eq.\,(\ref{eq7}) and make use of $
  	\ket{\psi_0(t)}=\int \psi_0(z,t)\ket{z}dz
$ to construct the optimal total Hamiltonian driving the process,
  \begin{widetext}
  	\begin{equation}
\begin{aligned}
	H(t)&=i \hbar (\ketz{\p_t\tilde{\psi}_0(t)}\braz{\tilde{\psi}_0(t)}-\ketz{\tilde{\psi}_0(t)}\braz{\p_t\tilde{\psi}_0(t)})\\
	&=i\hbar\int [\dot{\tilde{\psi}}_0(z,t) \tilde{\psi}(z^\prime,t)-\tilde{\psi}(z,t)\dot{\tilde{\psi}}(z^\prime,t)]\ket{z}\bra{z^\prime}dzdz^\prime\\
	&=i\int \eta m \omega_0 (z^\prime-z) \left( \f{m \omega_0 }{\pi \hbar e^{-2 \eta t \epsilon}}\right)^{1/2}(\mu+(z+z^\prime)e^{\eta t \epsilon}) \\&\times \exp\left(-\frac{m \omega_0\left(2 \mu^{2}+(z^{2}+{z^\prime} ^{2}) e^{2 \eta t \epsilon}+2 \mu(z+z^\prime) e^{\eta t \epsilon}\right)}{2 \hbar}-\eta t \epsilon \right)\ket{z}\bra{z^\prime}dz dz^\prime,
\end{aligned}
\end{equation}
  \end{widetext}
  where we have used the fact that $\ket{\psi_0(t)}=\ketz{\tilde{\psi}_0(t)}$. This is evident since
  \begin{equation}
  \begin{aligned}
  	\braketz{\dot{\psi}_0(t)}{\psi_0(t)} &= \int \f{\partial}{\partial z}(\alpha e^{- \beta(z+\gamma)^2}) \alpha e^{- \beta(z^\prime+\gamma)^2} \braket{z}{z^\prime} dz dz^\prime\\
  	&=\int -2 \alpha^2 \beta (z+\gamma)e^{-2\beta(z+\gamma)^2} dz =0
  \end{aligned}
  \end{equation}
  where $\alpha, \beta$ and $\gamma$ are coefficients that can be read off from Eq.\,(\ref{eq32}).
  
\section{V. Discussions}
It is worth noting that since $\bra{z}H(t)\ket{z^\prime}\neq0$ for $z\neq z^\prime$, the Hamiltonian we just solved for is non-local in space. One may wonder if it is possible to find a localized Hamiltonian that does the job, and it turns out that it is possible only in limited cases. To better understand, consider the Bohmian mechanics, where by writing an arbitrary wave function in the polar form $\psi=R \exp (i S / \hbar)$ and substituting it into the Schr\"{o}dinger equation $i \hbar \frac{\partial \psi}{\partial t}=(-\frac{\hbar^{2}}{2 m} \nabla^{2}+V) \psi$, one can obtain the continuity equation $\frac{\partial R}{\partial t}=-\frac{1}{2 m}(R \nabla^{2} S+2 \nabla R \cdot \nabla S)$ and the quantum Hamilton-Jacobi equation $\frac{\partial S}{\partial t}=-(\frac{|\nabla S|^{2}}{2 m}+V-\frac{\hbar^{2}}{2 m} \frac{\nabla^{2} R}{R})$ \cite{PhysRev.85.166}. Since the potential $V$ only appears in the latter, it can be easily solved,
\begin{equation}\label{eq35}
	V=-\f{\p S}{\p t}- \f{|\nabla S|^2}{2m}+\f{\hbar^2}{2m}\f{\nabla^2 R}{R}.
\end{equation}
Note that the potential $V$ obtained in this way will always be localized (i.e., $\bra{z}V\ket{z^\prime}=0$ for $z \neq z^\prime$) since we have assumed it to be so in the Schr\"{o}dinger equation $i \hbar \frac{\partial \psi}{\partial t}=(-\frac{\hbar^{2}}{2 m} \nabla^{2}+V) \psi$. However, such a solution of the potential $V$ is valid only if we also plug $R$ and $S$ into the continuity equation, which as a constraint restricts the class of possible solutions of the Hamiltonian given its assumed form. For the subclass of the wave functions where $R$ and $S$ satisfy the continuity equation, this suggests that we can always find a local potential $V$ that generates the dynamics described by the wave function $\psi$; for the subclass of the wave functions where $R$ and $S$ does \textit{not} satisfy the continuity equation, however, this implies that there does not exist a local potential generating the dynamics of the wave-function $\psi$.

To give a concrete example, consider the following 1-$d$ wave packet,
\begin{equation}
	\psi(z,t)=\left(\frac{m \omega}{\pi \hbar}\right)^{\frac{1}{4}} e^{-\frac{m \omega}{2 \hbar}(z-\mu t)^{2}} e^{i\left(\sqrt{m \hbar \omega} z-\frac{1}{2} \hbar \omega t\right) / \hbar}.
\end{equation}
By writing $\psi(z,t)$ in the polar form to obtain $R$ and $S$, and using the equation Eq.\,(\ref{eq35}), one can solve for $V$,
\begin{equation}
	V=-\frac{1}{2} \hbar\omega +\frac{1}{2} m \omega^{2}(z-\mu t)^{2}.
\end{equation}
Now plug $R$ and $S$ also into the continuity equation, which gives rise to the constraint $\mu=\sqrt{\hbar \omega/m}$. This means that for this particular wave-function $\psi(z, t)$, a localized driving potential is possible if and only if $\psi(z, t)$ travels at the speed $\mu=\sqrt{\hbar \omega /m}$, as any other choices of $\mu$ will break the continuity equation.

The above findings suggest that there exists a large class of wave functions that cannot be described by the Schr\"{o}dinger equation with a localized potential, even for a single particle system \cite{modanese2018time}. Nevertheless, they can always be described with a non-local Hamiltonian via Eq.\,(\ref{eq7}). This kind of non-local Hamiltonian does not have to be fundamental --- it could also represent an effective potential produced by course graining, or other degrees of freedom. It is intriguing whether those time-dependent wave functions whose existence requires a non-local Hamiltonian are prevalent in reality, since it seems that they are far more in number than those which can be described by a local Hamiltonian. A more careful study of the properties of those peculiar wave functions and their non-local Hamiltonians may reveal what might have been long overlooked in the study of quantum mechanics, and enable the exploitation of their possible advantages.

\section{VI. Conclusion}

Our result implies that any unitarily evolved density matrices, even for those that previously can only be well described by a master equation, can now be described by a time-dependent Hamiltonian (Eqs.\,(\ref{eq7}) and (\ref{eq23})). For instance, the formalism can be used to describe the continuous quantum measurement process and hence the continuous wave function collapse process \cite{hu2023describing}. It can be also used to calculated the effective Hamiltonian of a gate implementation via quantum tomography, from which one can compare how much the actual Hamiltonian is off from the designed Hamiltonian so as to correct, fine-tune and optimize the actual Hamiltonian used for quantum gates \cite{PhysRevLett.122.020504,PhysRevLett.124.160502,siva2022time}. Moreover, when the formalism is generalized for non-unitary evolutions, it leads to a brand new and unified framework to describe the open quantum system dynamics in all regimes \cite{hu2023probabilistic}, which states that \textit{any} continuous open quantum system dynamics can be regarded as the combined effects of a time-dependent Hamiltonian and probabilistic combinations of unitary operators. 

\section{VII. Acknowledgement}
We are grateful to Luiz Davidovich and Yakir Aharonov for helpful discussions. This work was supported by the Army Research Office (ARO) under Grant No. W911NF-22-1-0258.

\appendix
\nocite{*}
\bibliographystyle{ieeetr}
\bibliography{apssamp2}
\end{document}